\documentclass{article}
\usepackage{graphicx}
\usepackage{tikz}
\usetikzlibrary{shapes.geometric, arrows.meta, positioning, calc, shapes}
\usepackage[utf8]{inputenc}
\usepackage[margin=1in]{geometry} 
\usepackage{setspace}
\usepackage[backend=biber, style=apa]{biblatex}
\usepackage{booktabs}
\usepackage{hyperref}
\usepackage{amsmath}
\usepackage{float}
\usepackage{multicol} 
\usepackage{microtype} 
\usepackage{hyperref} 
\usepackage{cleveref}
\usepackage{tabularx} 
\usepackage{makecell} 
\usepackage{subcaption} 

\providecommand{\keywords}[1]{\textbf{Keywords:} #1}

\newcolumntype{C}{>{\centering\arraybackslash}X}

\title{SCAN: A Decision-Making Framework for Effective Task Allocation with Generative AI}

\author{
    Fendi Tsim$^{1*}$ \quad 
    Alina Gutoreva$^{2}$
}

\begin{document}

\maketitle

\begin{center}
{\small
$^{1}$Independent Researcher, London, United Kingdom \\
$^{2}$School of IT and Engineering, Kazakh-British Technical University, Almaty, Kazakhstan\\
$^{*}$Corresponding author. Email: \href{mailto:fenditsim@gmail.com}{fenditsim@gmail.com}\\[0.4em]
}
\end{center}

\renewcommand{\thefootnote}{\fnsymbol{footnote}}

\begin{abstract}
We introduce SCAN---a human-centric decision-making framework to facilitate learners for effective task allocation with Generative Artificial Intelligence (GenAI) based on Vygotsky’s Zone of Proximal Development and Metacognition. In SCAN, we systematize and formalize AI-human interaction by introducing a task-identification approach with four ``sub-zones'': Substitute, Complement, Aid, and Non-negotiable. After describing the four sub-zones, we demonstrate how SCAN framework can be applied for knowledge workers in the workplace and students in the education to metacognitively ''scan" their use of Generative AI. We then discuss how such framework can be related to cognitive load theory, cognitive offloading, sycophancy, three decision making modes in human-AI interactions (automation, augmentation, and collaboration), future of work such as upskilling and deskilling, and how it accounts for both human-human and human-AI learning. We propose that SCAN offers a great starting point before discussing whether GenAI complements or replaces our abilities when completing a task, with a general objective of sustaining lifelong learning, and a specific goal of reaching hybrid intelligence. 
\end{abstract}

\keywords{Human-AI interaction, Generative AI, Hybrid Intelligence, Metacognition, AI-Assisted Decision Making}

\section{Introduction}

How do we complete a task with assistance? For instance, Annie, who is more senior than Ben in the company, assigns Ben a task to complete. Ben would, of course, begin by identifying it: clarifying the scope (What is this task about?), the objectives (What are the objectives that need to be met?), and what a successful output looks like (What does the outcome look like?). Before proceeding, he confronts to a quieter, more personal question: \emph{Do I actually know how to do this?} He has written reports before, but this one requires a methodology he has only read about. He knows enough to recognize the gap, yet not enough to close it alone. This moment of \emph{self-assessment}---locating himself between what he can do independently, and what remains just out of reach---is where the real work of task assignment begins. With Annie's guidance bridging that gap, Ben completes the report, and in doing so, gains new or consolidates existing skill and thus grows professionally. Now, suppose Ben has access to a powerful GenAI tool instead. The three-stage cycle (task, process, and evaluation identification) remains, but each stage multiplies into a decision: \emph{Should I let GenAI draft the structure, or do I do that myself? Can I trust its data retrieval? Should I verify its conclusions?} Without first asking what he knows and does not know, Ben cannot answer any of these questions well. He might end up either offloading too much, which in turn erodes the very skills the task was meant to develop, or too little---using an able tool as a mere spell-checker. As a result, the problem is not about GenAI's capability; it is Ben's not understanding how to use AI in the task \emph{effectively}.

The recent rise of GenAI has made ineffective task allocation a widespread concern. GenAI offers ``general knowledge'' of a task based on its ability to collect, filter, and synthesize general information based on the data that it is trained on, as well as information from the Internet \parencite{feuerriegel2024generative}. In task completion, GenAI can be a great ``scaffold'' \parencite{ngu2025generative} to augment our capabilities at task, and thus potentially leading to substantial benefits such reducing time for task completions, and allocating unused cognitive resources to other tasks. Yet, empirical research report that GenAI assistance leads to a series of issues such as over- and under-reliance \parencite{klingbeil2024trust, de2023psychological}, automation bias \parencite{bangerl2025creaitive, darvishi2024impact}, critical thinking \parencite{tian2025learners}, and harming an individual's learning \parencite{bastani2025generative}, reducing one's intrinsic motivation when collaborating with GenAI \parencite{wu2025human}, and even reduction in one's metacognitive ability \parencite{fernandes2026ai}. These issues, we believe, share a common root: when individuals do not first assess their own knowledge relative to a task, they cannot determine where their competence ends, and more importantly, where AI assistance genuinely adds leverage. The delegation decision becomes arbitrary, and overreliance fills the vacuum. Most importantly, with the goal of reaching an equilibrium that maintains or improves our capabilities while leveraging GenAI's, how do we assign a task with GenAI effectively, i.e., for automating some of the processes, augmenting on what we cannot complete independently \parencite{hermann2025genai}? 

Existing approaches offer partial guidance. AI-centric solutions propose that the system itself determines task assignment based on modeled human and AI capabilities \parencite{fugener2025roles}, but recent evidence raises concerns about the resulting loss of agency and responsibility \parencite{salatino2025influence, vallor2024find}. Human-centric heuristics, such as \citeauthor{mollick2024co}'s (\citeyear{mollick2024co}) distinction between ``Just Me,'' ``Delegated,'' and ``Automated'' tasks, are more practically grounded but describe outcomes of the assignment decision rather than providing a procedure rooted in the learner's knowledge state. A growing body of research examines this further in the education by incorporating psychological perspective \parencite{sidorkin2025leapfrogging, ganuthula2025artificial, atchley2024human}. For instance, \textcite{nasr2025exploring} reports that the effective use of GenAI in education depends on the quality of human–AI interaction. Yet, what determines interaction quality at the individual task level, and how an individual's own knowledge state should inform it, remains underspecified.

In this paper, we introduce SCAN, a human-centric decision-making framework that gives learners a systematic procedure for assigning tasks with GenAI. SCAN consists of two structural components in psychology: \citeauthor{Vygotsky1978}'s (\citeyear{Vygotsky1978}) work of Zone of Proximal Development (henceforth \emph{ZPD}) and Metacognition \parencite{flavell1979metacognition}. Together, they identify where AI belongs in a given task ---in the learner's real-time awareness of what she knows, what she does not, and where external assistance adds genuine leverage. Extending ZPD with a ``known to GenAI'' zone \parencite{yan2024promises}, SCAN identifies four sub-zones for task assignment: Substitute, Complement, Aid, and Non-negotiable. Returning to Ben, SCAN gives him a concrete answer: tasks he cannot perform at all belong in Substitute; tasks where his knowledge and GenAI's general capability overlap productively belong in Aid; tasks where he already holds the expertise, and simply delegates execution belong in Complement; and tasks requiring human judgment, accountability, or mentorship from another person remain Non-negotiable regardless of what GenAI can do. The overarching goal of SCAN is \emph{sustained learning} with GenAI---preserving the conditions under which Ben continues to grow rather than quietly deskilling \parencite{simkute2025ironies}. 

In what follows, we begin by presenting the foundation of SCAN including its four core components. We then demonstrate how SCAN can be applied in two common real-life scenarios for individuals interacting with GenAI, where knowledge workers or learners, who always have domain-specific knowledge before the rise of GenAI, and those who are currently accumulating knowledge during the GenAI era. We then discuss how SCAN can be related to cognitive offloading and sycophancy, three decision-making modes in human-AI interactions (automation, augmentation, and collaboration), as well as how it can be evolved as upskilling \parencite{lira2025learning} and deskilling \parencite{simkute2025ironies} over time. As each of us learns moment by moment in the GenAI era, and human shall remain in the loop when interacting with GenAI, we propose SCAN offers a starting point, with the goal of reaching ``Hybrid Intelligence'' \parencite{dell2023navigating}.

\section{Theoretical Framework}
\subsection{Task Paradigm}
In this section, we introduce SCAN's Task Paradigm---one of the SCAN's structural components that illustrates an individual's learning and development via collaborative and peer learning based on \citeauthor{Vygotsky1978}'s (\citeyear{Vygotsky1978}) work of Zone of Proximal Development (ZPD). ZPD describes a scenario when an individual (e.g., a learner) requires to complete a task at hand: she can complete it independently or not, as well as when she can complete it with guidance and support from others who are more knowledgeable than her such as teachers or experts as a contingent support \parencite{van2019scaffolding}. Building on the notion of learning precedes development as a \emph{dynamic}, rather than a static, assessment \parencite{allal2000assessment}, his theory can be represented as three zones (or levels) of development. 

First, the \emph{Zone of Actual Development}, or the ``actual developmental level'', is where a learner can complete a task independently with her own knowledge, without any assistance of others. According to \citeauthor{Vygotsky1978}'s (\citeyear{Vygotsky1978}), it is whereby a person's mental functions have been established as a result of certain already completed developmental cycles. Central to this zone is the notion of \emph{task-specific knowledge}: a learner's self-recognized knowledge that is directly applicable to the task at hand. This type of knowledge is a subset of her broader domain-specific knowledge \parencite{alexander1988interaction, tricot2014domain, devine1995domain}. For instance, when we purchase a cup of coffee in a cafe, we need to know, and thus tell a bartender, what to order (task-specific), rather than the ratio between milk and coffee beans (domain-specific). In what follows, we label this zone as ``known to learner'', implying that a learner has task-specific knowledge about a task she needs to complete. 

Second, the \emph{Zone of Potential Development} is where a learner cannot complete a task independently, even with the assistance of More Knowledgeable Others (MKO). If a learner needs to complete a task in this zone, this overloads her working memory, affecting her performance and learning ability. We label this zone as ``unknown to learner'', implying that she has no task-specific knowledge for task completion. 

Finally, the \emph{Zone of Proximal Development}, or ZPD, is where an individual's learning behavior occurs via her social interactions with others. It captures what a learner can accomplish with MKO, beyond what she can achieve independently but within the bounds of what is currently learnable \parencite{Vygotsky1978, clara2016}. In other words, although a learner has some task-specific knowledge in this zone, it is not sufficient for independent task completion. Therefore, she requires the assistance of more knowledgeable others as ``scaffold'' \parencite{wood1976role, van2010scaffolding}, through learning with peers \parencite{tenenbaum2020effective} and cooperative learning \parencite{erbil2020review}.

\begin{figure}[htbp]
    \centering
    
    \begin{subfigure}[b]{0.38\textwidth}
        \centering
        \begin{tikzpicture}[scale=0.85]
            \def\innerRadius{1}
            \def\middleRadius{1.75}
            \def\outerRadius{2.5}
            
            \useasboundingbox (-2.7, -2.7) rectangle (2.7, 2.7);
            
            \fill[white] (0,0) circle (\outerRadius);
            \fill[gray!30] (0,0) circle (\middleRadius);
            \fill[black] (0,0) circle (\innerRadius);
            
            \draw[thick] (0,0) circle (\innerRadius);
            \draw[thick] (0,0) circle (\middleRadius);
            \draw[thick] (0,0) circle (\outerRadius);
            
            \node[white, font=\scriptsize] at (0,0) {\shortstack{Known to\\learner}};
            
            \node[font=\small] at (0,1.3) {ZPD};
            
            \node[font=\scriptsize] at (0,2.05) {\shortstack{Unknown to\\learner}};
        \end{tikzpicture}
        \caption{Concentric circles model}
        \label{fig:concir}
    \end{subfigure}
    \hfill 
    \begin{subfigure}[b]{0.58\textwidth}
        \centering
        \begin{tikzpicture}[scale=0.9] 
            \useasboundingbox (-5.5, -2.2) rectangle (5.5, 2.2);
            
            \coordinate (V2L) at (-3.6, 0);
            \coordinate (V2R) at (-2.2, 0);
            \def\radTwo{1.1}
            
            \begin{scope}
                \clip (V2L) circle (\radTwo);
                \fill[gray!30] (V2R) circle (\radTwo);
            \end{scope}
            
            \draw[thick] (V2L) circle (\radTwo);
            \draw[thick] (V2R) circle (\radTwo);
            
            \node[font=\tiny] at (-4, 0) {\shortstack{Known to\\learner}};
            \node[font=\tiny] at (-1.8, 0) {\shortstack{Unknown\\to learner}};
            \node[font=\scriptsize] at (-2.9, 0) {ZPD};
            
            \draw[->, ultra thick, >=stealth] (-0.5, 0) -- (0.5, 0);
            
            \coordinate (V3L) at (2.25, 0.45);   
            \coordinate (V3R) at (3.75, 0.45);   
            \coordinate (V3B) at (3.0, -0.6);    
            \def\radThree{1.2}
            
            \begin{scope}
                \clip (V3L) circle (\radThree);
                \clip (V3R) circle (\radThree);
                \fill[gray!30] (V3R) circle (\radThree);
            \end{scope}
            
            \begin{scope}
                \clip (V3L) circle (\radThree);
                \clip (V3R) circle (\radThree);
                \fill[gray!30] (V3B) circle (\radThree);
            \end{scope}
            
            \draw[thick] (V3L) circle (\radThree);
            \draw[thick] (V3R) circle (\radThree);
            \draw[thick] (V3B) circle (\radThree);
            
            \node[font=\tiny] at (1.8, 0.7) {\shortstack{Known to\\learner}};
            \node[font=\tiny] at (4.2, 0.7) {\shortstack{Unknown\\to learner}};
            \node[font=\tiny] at (3.0, -1.2) {\shortstack{Known to\\GenAI}};
            
            \node[font=\scriptsize] at (3.0, 0.85) {N};
            \node[font=\scriptsize] at (3.0, 0.12) {A};
            \node[font=\scriptsize] at (2.3, -0.2) {C};
            \node[font=\scriptsize] at (3.7, -0.2) {S};
            
        \end{tikzpicture}
        \caption{From ZPD to SCAN's Task Paradigm}
        \label{fig:venn_evolution}
    \end{subfigure}

    \caption{Visualizations of ZPD (Concentric Circles and Venn Diagram) and SCAN's Task Paradigm}
    \label{fig:zpd_models}
\end{figure}
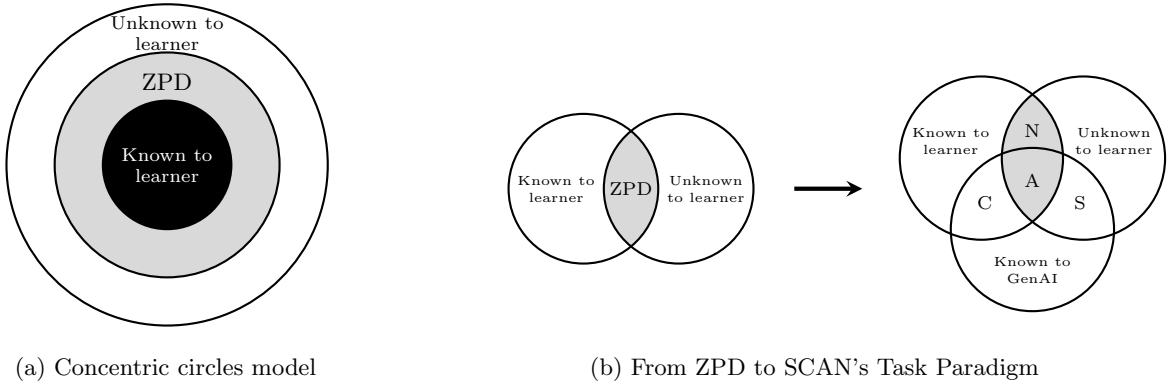

To the best of our knowledge, there are two common visual representations to illustrate its dynamics: concentric circles and a Venn Diagram. The former is the three nested circles with the Zone of Actual, Proximal, and Potential Development as the inner, middle, and outer circles respectively (\cref{fig:concir}). As ZPD is the middle circle, it highlights the learner's expandability of the inner circle towards the outer circle. The second form of ZPD's visualization is a Venn Diagram with two circles: the Zone of Actual and Potential Development are left and right circles respectively (\cref{fig:venn_evolution}; left). The ZPD is the overlapping, shaded area of two circles, which illustrates how the level of task-specific knowledge varies across three zones of development. In what follows, we visualize ZPD with the Venn Diagram, as it clearly delineates the boundary within which a learner may seek assistance from MKO or GenAI---a distinction central to SCAN that the concentric circles representation fails to capture. Such boundary, as we shall see, is essential in exploring some popular concepts in human-AI interactions when considering how GenAI can affect a learner while completing a task, such as augmentation and automation.

A vast body of literature examines how technology can be integrated with ZPD to support learning \parencite{crook1991computers, sun2023systematic, hrastinski2009theory, zuo2023effects, sharma2021information, hobert2023chatbots}, with recent work extending this to GenAI in higher education \parencite{cai2025exploring, ferguson2022ai}. Complementing this body of literature, we propose adding a new circle (or zone) called ``known to GenAI'' in the ZPD, constructing the foundation of the SCAN framework. It is assumed that GenAI offers ``general knowledge'' of a task based on its ability to collect, filter, and synthesize general information based on the data is trained on, as well as information from the Internet \parencite{feuerriegel2024generative}. This addition in ZPD forms the \emph{task paradigm} of SCAN---a map with the four sub-zones located within Vygotsky's three zones of development (\cref{fig:venn_evolution}; right), which are denoted as Substitute (S), Complement (C), Aid (A), and Non-negotiable (N).

Adding ``known to GenAI'' in the ZPD yields two essential benefits. First, it can separate GenAI from human experts in MKO as ``a digital scaffold'' \parencite{chien2025learning} to assist a learner. This can, for instance, highlight the struggles educators face when adopting GenAI in their work related to provide assessments, teaching, and learning delivery in the education setting \parencite{ogunleye2024systematic, sidorkin2025leapfrogging}. Secondly, and more importantly, it offers a clear direction, both theoretically and visually, to further examine when a learner can leverage GenAI to augment her existing knowledge, or from human experts, of a given task. 

In what follows, we are going to describe each sub-zone in the task paradigm in detail. The first two sub-zones are Substitute and Complement, which correspond to ``replacement'' and ``complementary''---two terms that have been widely investigated in understanding GenAI-assisted decision making \parencite{loaiza2024epoch, hemmer2025complementarity, makela2024complement}.

\subsubsection{Substitute (S)}
Substitute as a concept has been widely discussed since the rise of GenAI, with a particular focus on human automating tasks and the risk being replaced \parencite{simkute2025ironies, feuerriegel2024generative}. In SCAN, it is at the intersection between Vygotsky's Zone of Potential Development and Known to GenAI, implying that a learner has no task specific knowledge to complete a task, even with the help of more knowledgeable others. However, GenAI has general knowledge that could complete such task with a generic approach, rather than a specific one. For instance, GenAI knows how to process a large amount of data and perform complex computations beyond human capabilities \parencite{chiriatti2024case}.

Automation occurs when a learner completes a task in this sub-zone. She is prone to cognitive offloading \parencite{shukla2025skilling, frey2017future}: she substitutes her cognitive ability with GenAI's ability when completing a task. This could result in severe consequences on her cognitive load, performance \parencite{stadler2024cognitive} and learning ability if she needs to complete it. Apart from that, she is prone to automation bias \parencite{romeo2025exploring}: she might send a prompt with a few or none inputs (e.g., "Write me a code for..."), and adopt GenAI's response accordingly.

\subsubsection{Complement (C)}
In contrast to Substitute, Complement is at the intersection between Vygotsky’s Zone of Actual Development and Known to GenAI, implying that a learner has \emph{sufficient} task specific knowledge to complete independently, and verify the output of, a task, and GenAI has general knowledge. The nature and structure when completing a task in this zone alters: instead of completing a task on her own, a learner has ``an option'' to assign it to GenAI, and thus monitor the output. Although choosing the latter option indicates more sub-tasks she needs to complete such as verifying the GenAI output, it frames human and GenAI as a team \parencite{schmutz2024ai}, thus leading to substantial benefits such reducing time for task completions, and allocating unused cognitive resources to other tasks, reaching ``Human-AI collaboration'' \parencite{ganuthula2025artificial, sundar2020rise}.

Recent studies examine ``Human-AI Complementarity'' both theoretically and empirically \parencite{vaccaro2024combinations, hemmer2025complementarity}. The empirical finding shows that human-AI complementarity, or human-AI synergy has been observed rarely (a recent observation is from \textcite{riedl2025quantifying}), suggesting the need to revise its definition and assessment. SCAN shows that achieving such collaboration can be task dependent, and proposes that such dependency is largely attributed to whether a learner has task specific knowledge beforehand. So, for instance, an experienced researcher would conduct a literature search quicker and more comprehensive than a novice researcher when using a GenAI-assisted software for literature search (e.g., Google's Scholars Tab).\\

\noindent
Next, we look at two other sub-zones of SCAN that are located at the ZPD, which differ in the source of a scaffold that a learner augments her capabilities with (more knowledgeable others or GenAI). We highlight that by having two separate sub-zones in the ZPD that differ in source of assistance, we do not reject the possibility that more knowledge others would use GenAI to assist (for instance, a teacher might build an interactive visualization with GenAI for teaching an abstract concept in Psychology).

\subsubsection{Aid (A)}
Locating at the intersection between ZPD and Known to GenAI, Aid denotes the occurrence of GenAI as a scaffold for a learner to complete a task \parencite{ngu2025generative, reicherts2025ai}. A learner has some but insufficient task specific knowledge to complete the task independently unless assistance from GenAI. GenAI can be a learner's scaffold to complete  by providing general knowledge of a task (which a learner lacks). In other words, a learner can enhance her capabilities by leveraging GenAI's, which is a sign of human-GenAI augmentation (completing a task with GenAI performs better than without it). For instance, a learner can enhance her critical thinking ability when interacting with GenAI \parencite{drosos2025makes, guo2025critical}. The type of a task matters a lot for the existence of such augmentation: when the task requires creativity, human-AI augmentation appears \parencite{huang2025unlocking}. If, instead, it is a decision task, which human decision maker already complete most of the part of a task, GenAI is mostly treated as a supplement, leading no augmentation \parencite{vaccaro2024combinations}. 
Essentially, SCAN facilitates self-regulated learning.

Aid differs from Substitute. In the latter, a learner has (1) no knowledge about the task, and (2) cannot complete the task independently. Suppose an individual needs to create an interactive dashboard with a programming language, or known as Vibe Coding \parencite{sarkar2025vibe}. If she does not possess any knowledge about how to program an interactive dashboard, she would prefer to offload it to GenAI (automation), leading to Substitute. If, instead, she possesses some knowledge about it (e.g., knowing how to create a chart, but not how to build a dashboard with relevant packages and functions), and identifies this task in the Aid zone, GenAI can assist her thinking on how to build such dashboard \parencite{reicherts2025ai}.

\subsubsection{Non-negotiable (N)}
Non-negotiable aligns with the original definition of ZPD \parencite{Vygotsky1978}. Unlike S, C, and A, which extend ZPD by introducing GenAI in the development space, this sub-zone retains the original human-human structure that Vygotsky described: a learner completes a task with guidance and support from a more knowledgeable other (MKO) such as a supervisor, mentor, expert, or teacher, rather than with GenAI. Examples are mentorship relationships, personnel decisions, and ethical reviews where professional judgment, accountability, and relational attunement are central.

Apart from the source of scaffold, there are two fundamental differences between Aid and Non-negotiable. First, the type of knowledge required for task completion differs. In Aid, the general knowledge from GenAI is sufficient to bridge the gap between what a learner knows and what a task requires. In Non-negotiable, however, the knowledge required cannot be generalized. Rather, it is tacit, situated, and accumulated through sustained personal experience that resists codification \parencite{polanyi1966tacit}. For instance, a mentor offers career guidance within a specific organizational culture, or a clinician exercises judgment at a patient's bedside. In these tasks, the required knowledge are both context-specific and more importantly, inseparable from the relational and experiential conditions in which it was formed. 

Secondly, what the scaffold is designed to produce in the learner afterward differs as well. In Aid, GenAI is mainly epistemic by providing general knowledge the learner lacks, enabling task completion. In contrast, the scaffold's role in Non-negotiable is relational and developmental. On top of providing knowledge, the human MKO provides a participatory context in which cognitive \parencite{collins1989cognitive}, behavioral \parencite{bandura1977social}, and social development \parencite{lave1991situated} occurs. For instance, the MKO models professional judgment, ways of reasoning, and patterns of engagement with a domain that the learner gradually internalizes \parencite{Vygotsky1978, wood1976role}. Thus, the learner in Non-negotiable acquires professional identity, relational competence and human capabilities such as empathy, situated judgment, and professional presence that GenAI structurally cannot replicate \parencite{loaiza2024epoch}; difficult to be automated \parencite{WEF2025NewEconomySkills}; domains where a learner's identity is threatened and when evaluative criteria are ambiguous \parencite{morewedge2022preference}; or where accountability must remain traceable to a human agent \parencite{vallor2024find}. In short, these are essential dimensions of development that GenAI cannot produce: delegating such a task to GenAI would short-circuit the developmental process entirely.

\begin{figure}[H]
\begin{center}
\makebox[\textwidth][c]{
\begin{tikzpicture}[
    scale=0.75, transform shape,
    node distance=1.5cm and 1.2cm,
    base/.style = {rectangle, draw, thick, align=center, font=\sffamily\small},
    startnode/.style = {base, fill=orange!5, minimum width=5cm, minimum height=1.2cm},
    decision/.style = {diamond, draw, thick, fill=white, aspect=2.5, inner sep=0pt, font=\sffamily\bfseries\small, text width=3.5cm, align=center},
    zoneStyle/.style = {base, draw=gray!80, fill=gray!10, text width=4cm, minimum height=2cm, font=\sffamily},
    cornerLabel/.style = {fill=gray!70, text=white, font=\sffamily\bfseries\tiny, inner sep=2pt, minimum width=0.6cm, minimum height=0.4cm},
    arrow/.style = {thick, -{Stealth[scale=1.2]}},
    bypass/.style = {thick, -{Stealth[scale=1.2]}, color=black!90},
    loopArrow/.style = {ultra thick, dashed, gray!60, -{Stealth[scale=1.5]}}
]

    \node (start) [startnode, minimum width=14cm] {Metacognitive Knowledge \& Monitoring (Real-time Task Evaluation)};

    \node (dec1) [decision, below=1.2cm of start] {Do I possess\\ task-specific knowledge?};

    \node (dec2) [decision, below=2.8cm of dec1] {Is GenAI\\ involvement needed?};

    \node (sub) [zoneStyle, left=of dec1] {\textbf{Substitute (S)}\\[0.5em] Automation};
    \node[cornerLabel, anchor=north west] at (sub.north west) {S};

    \node (comp) [zoneStyle, right=of dec1] {\textbf{Complement (C)}\\[0.5em] Collaboration};
    \node[cornerLabel, anchor=north west] at (comp.north west) {C};

    \node (aid) [zoneStyle, left=of dec2] {\textbf{Aid (A)}\\[0.5em] Augmentation with GenAI};
    \node[cornerLabel, anchor=north west] at (aid.north west) {A};

    \node (non) [zoneStyle, right=of dec2] {\textbf{Non-negotiable (N)}\\[0.5em] Augmentation with human experts};
    \node[cornerLabel, anchor=north west] at (non.north west) {N};

    \node (feedback) [startnode, below=2.8cm of dec2, minimum width=14cm] {Metacognitive Control (Reflection)};

    \draw [arrow] (start) -- (dec1);
    
    \draw [arrow] (dec1) -- node[above, font=\sffamily\small] {No} (sub);
    \draw [arrow] (dec1) -- node[above, font=\sffamily\small] {Sufficient} (comp);
    \draw [arrow] (dec1) -- node[right, font=\sffamily\small] {Some but insufficient} (dec2);

    \draw [arrow] (dec2) -- node[above, font=\sffamily\small] {Yes} (aid);
    \draw [arrow] (dec2) -- node[above, font=\sffamily\small] {No} (non);

    \draw [bypass] (sub.west) .. controls ($(sub.west)+(-3.5, 0)$) and ($(feedback.north west)+(-3.5, 3)$) .. ($(feedback.north west)+(0.3,0)$);
    \draw [bypass] (comp.east) .. controls ($(comp.east)+(3.5, 0)$) and ($(feedback.north east)+(3.5, 3)$) .. ($(feedback.north east)-(0.3,0)$);

    \draw [arrow] (aid.south) -- (aid.south |- feedback.north);
    \draw [arrow] (non.south) -- (non.south |- feedback.north);

    \draw [loopArrow] (feedback.west) to [out=150, in=180, looseness=1.8] node[pos=0.5, right=4pt, font=\sffamily\small] {Learning} (start.west);

\end{tikzpicture}
}
\end{center}
\caption{SCAN's step-by-step Implementation Visualization}
\label{fig:illustration}
\end{figure}
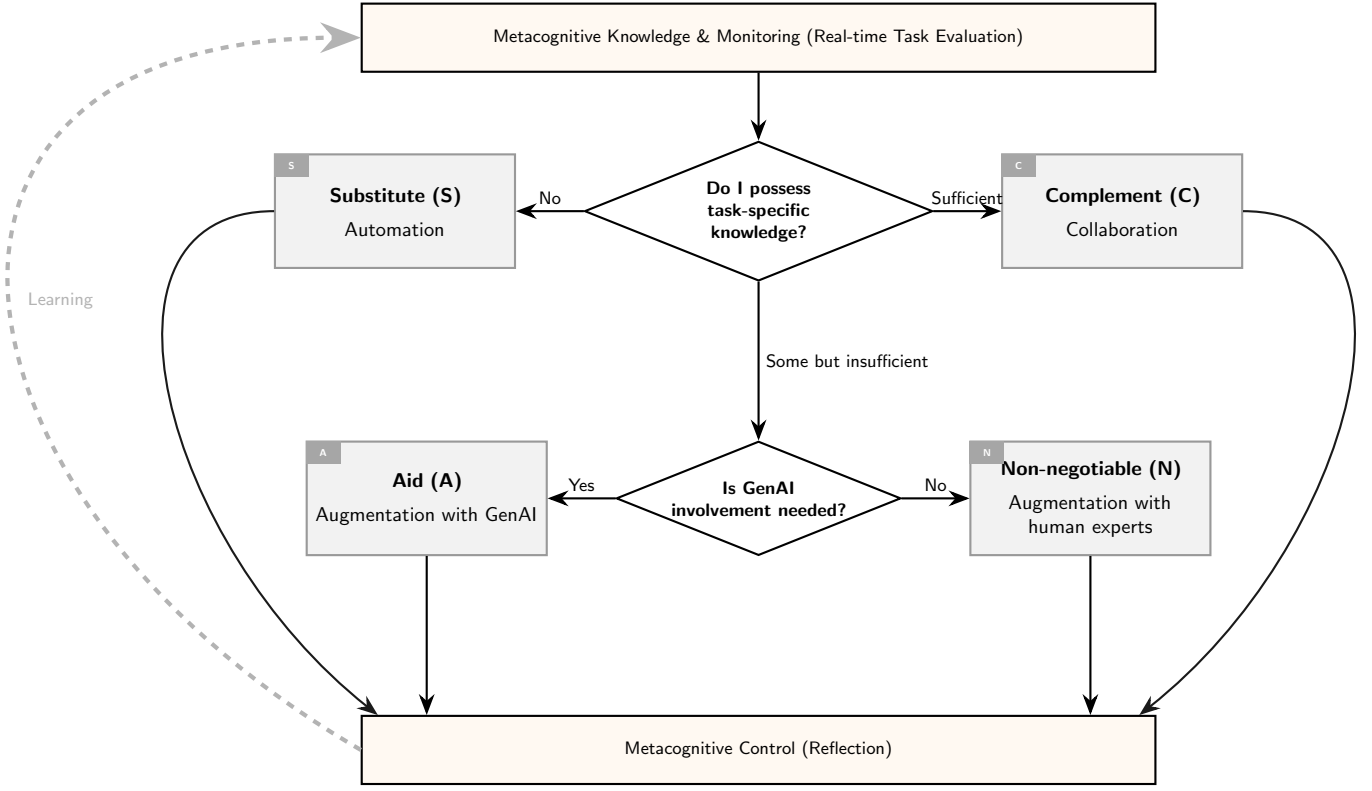

\subsection{Metacognition}
 
Recent work on metacognition in human--AI interactions reveals a consistent pattern: as reliance on generative AI increases, accuracy in detecting AI errors decreases, yet confidence does not \parencite{fernandes2026ai,kasneci2023chatgpt}. This performance--metacognition dissociation allows the individuals to outsource cognitive generation while nominally retaining evaluative responsibility, but their capacity to exercise that responsibility is degraded by the act of delegation itself. Critically, this is not resolved by greater AI literacy, which \textcite{fernandes2026ai} found to be associated with \emph{lower} metacognitive accuracy. Experimental evidence further shows that users overrely on AI advice even when it contradicts both available context and their own prior judgment \parencite{klingbeil2024trust}. Metacognition in AI environments is, therefore, not purely internal but distributed across the human--AI system, and is actively distorted by it. Within SCAN's task paradigm, this positions metacognition as the control layer governing whether cognitive work is delegated, monitored, or retained, and directly motivates the zone architecture of the previous section between a learner and GenAI: Substitute, Aid, and Complement differ precisely in the metacognitive demand they impose. Analogous to a traditional mindfulness body scan meditation, which systematically directs non-judgmental, focused awareness to distinct physical regions to evaluate baseline state and alignment \parencite{carmody2008mindfulness}, the SCAN framework operationalizes a literal ``cognitive scan'' of one's workload. In this reflective state, the learner dynamically scans a pending task's structural constraints against their own immediate competence boundaries and available machine capabilities. 

Metacognition, defined as awareness and regulation of one's own cognitive processes \parencite{flavell1979metacognition, cox2005metacognition, tankelevitch2024metacognitive}, constitutes the second structural component of SCAN and its primary operational mechanism. Whereas SCAN's task paradigm delineates \emph{where} a task lies relative to a learner's competence boundary, metacognition governs \emph{how} that boundary is perceived, evaluated, and recalibrated in real time. Without sufficient metacognitive resolution, task classification collapses: learners misestimate their knowledge state and misallocate across sub-zones, an empirically documented failure mode in calibration error, overconfidence, and metacognitive bias. In human--AI interaction the stakes amplify: effective use of generative AI requires not only domain competence, but the capacity to evaluate the epistemic reliability and failure modes of AI outputs against one's own reasoning \parencite{kasneci2023chatgpt, bender2021stochastic, bergamaschi2025fast}. Without this, users default to over-reliance or under-utilization, both of which degrade performance and learning.

Classically, metacognition decomposes into knowledge, monitoring, and control \parencite{flavell1979metacognition, schraw1995metacognitive, nelson1990metamemory}. Knowledge encodes beliefs about task demands and personal competence, underwriting zone identification. Monitoring tracks performance and uncertainty in real time, enabling detection of zone misalignment. Control implements strategic adjustments, including whether to offload to AI, escalate to human expertise, or persist independently. Within SCAN these map onto task placement, real-time reassignment, and adaptive delegation, transforming SCAN from a static taxonomy into a self-correcting system grounded in measurable mechanisms. The verification bottleneck documented by \textcite{fernandes2026ai} and \textcite{klingbeil2024trust} is, in SCAN terms,a failure of monitoring and control: users place tasks in Complement or Aid but engage in Substitute-level oversight.

\subsubsection{Metacognitive Knowledge} 
Metacognitive knowledge refers to a learner’s structural beliefs and representations regarding their own cognitive agency, task parameters, and available strategic options \parencite{flavell1979metacognition, schraw1998metacognition}. This foundational knowledge requires \emph{learning}, that is, actively acquires cognitive architecture that develops progressively through structured training, explicit feedback, and iterative loops of self-regulated practice \parencite{dignath2008how, zimmerman2002becoming}. Within the SCAN paradigm, this construct serves as the critical representational structural precondition for initial zone identification—determining whether an upcoming task is partitioned into Substitute, Aid, or Complement. 
Contemporary accounts extend this to include calibrated self-models of competence, task demands, and strategy effectiveness, treating it as an internal, structured representation rather than a static belief system \parencite{fleur2021metacognition, popandopulo2023assessment}. Within the SCAN's task paradigm, this construct serves as the critical representational substrate and structural precondition for initial zone identification—determining whether an upcoming task is partitioned into Substitute, Aid, or Complement. 

In AI-mediated environments, this tendency is exacerbated by the surface
fluency of large language model outputs, which novices frequently conflate
with epistemic reliability \parencite{colville2026trust, safarov2026inaiwe}. This problem is compounded when AI operates as an information retrieval intermediary: algorithmically curated environments actively shape what content learners access and how they interpret it, requiring metacognitive knowledge to extend beyond self-assessment into calibrated evaluation of the search
environment itself \parencite{allen2025multidimensional,vu2000metacognitive}. 

This creates a tendency to overgeneralize AI competence across domains, frequently conflating surface-level linguistic fluency with genuine epistemic reliability \parencite{kasneci2023chatgpt}. Consequently, miscalibrated metacognitive knowledge drives systemic allocation errors, leading users to delegate high-level tasks requiring human judgment while manually retaining low-level tasks ripe for automation. Empirically, this miscalibration manifests as a breakdown in meaningful operational control; for instance, \textcite{kandul2023human} demonstrate significant disparities in how humans project predictability and evaluate machine boundaries during real--effort task distribution.

Crucially, this classification framework is structurally anticipatory, relying on domain schemas and prior experience to select cognitive policies under resource constraints \parencite{liu2025metacognitive, abdelshiheed2023metacognition}. This dependency explains why achieving effective human-AI coordination is fundamentally task- and expertise-dependent. For example, an experienced researcher leverages domain-specific schemas to rapidly filter and cross-examine GenAI literature search outputs, whereas a novice lacks the metacognitive grounding to detect subtle failures in machine reliability. Thus, improving descriptive AI literacy alone is insufficient to prevent misallocation; adaptive control requires continuous, latent calibration of these internal models. \textcite{colville2026trust} highlight that optimizing human-AI reliance depends on establishing this informed trust/distrust dynamic, while \textcite{safarov2026inaiwe} demonstrate that the interaction between explainable GenAI and foundational user expertise ultimately dictates whether a tool functions as a true cognitive complement or an unmonitored substitute.
 
\subsubsection{Metacognitive Monitoring} 
Metacognitive monitoring refers to the real-time, ongoing evaluation by which learners track their internal comprehension states, level of uncertainty, and performance trajectories during task execution \parencite{flavell1979metacognition, nelson1990metamemory}. Operating as the cognitive sensing layer, it functions as a closed-loop feedback mechanism that detects misalignments between intended and actual execution states. Within the SCAN framework, monitoring, or \emph{real-time evaluation}, provides the empirical foundation for \emph{real-time reassignment}, tracking whether an ongoing task remains stable within its designated zone or requires an immediate structural shift across the Substitute, Aid, or Complement boundaries as individual uncertainty or machine reliability fluctuates.

In generative AI-mediated environments, this monitoring architecture becomes a distributed, bidirectional process where users must simultaneously track personal cognitive limits and evaluate the epistemic reliability of machine outputs \parencite{kasneci2023chatgpt}. This process is highly susceptible to a fluency–accuracy dissociation: the stylistic surface fluency of large language models masks latent errors or hallucinations, severely reducing a user's error-detection sensitivity and causing a profound "monitoring illusion" or overconfidence effect \parencite{colville2026trust}. Such monitoring failures are often more predictive of suboptimal task performance than a native deficit in baseline domain knowledge itself. Users operating with impaired monitoring resolution fail to appropriately calibrate trust, sliding into debilitating automation bias or algorithmic aversion \parencite{dietvorst2015algorithm}. 

Empirically, this manifest breakdown in tracking performance and dynamic calibration highlights why human control must be reassessed during task distribution. As demonstrated by \textcite{kandul2023human}, users frequently struggle to accurately predict machine failure boundaries relative to human execution vectors in real-effort dynamics, which directly compromises human--AI task allocation. Furthermore, \textcite{schemmer2023should} emphasize that optimal reliance cannot depend on static or generalized trust; rather, it requires a task-specific, dynamic model of reliance calibration that adjusts in real time to shifting context. Within SCAN, high-resolution monitoring ensures fluid zone sensitivity, allowing a learner to recognize exactly when a task must transition—such as escalating a task from Complement to Aid when machine outputs conflict with context, or offloading to Substitute when manual cognitive resources are exhausted.
 
\subsubsection{Metacognitive Control} 
Metacognitive control represents the executive mechanism through which learners act upon monitoring data to select, adjust, or abandon ongoing cognitive strategies \parencite{flavell1979metacognition, pintrich2000role}. Within the SCAN framework, control operates at an analytical and retrospective temporal scale, enabling users to periodically update their internal assumptions regarding task difficulty, machine constraints, and optimal labor division. Beyond immediate task execution, control governs error attribution—empowering users to systematically differentiate between execution breakdowns caused by native baseline misunderstandings, inadequate prompt constraints, or miscalibrated trust in generative outputs \parencite{lee2004trust, koriat2007metacognition}. 

At a systemic level, metacognitive control serves as the primary engine for \emph{zone migration}. As domain-specific competence grows or model behavior patterns become clearer, a well-regulated learner actively adjusts their assignment policies, smoothly shifting recurrent workloads from Substitute toward Aid or Complement \parencite{hoff2015trust}. At a systemic level, metacognitive control serves as the primary engine for \emph{zone migration}. As domain-specific competence grows or model behavior patterns become clearer, a well-regulated learner actively adjusts their assignment policies, smoothly shifting recurrent workloads from Substitute toward Aid or Complement \parencite{hoff2015trust}. Notably, the motivational
constructs underlying this migration---goal orientation, self-efficacy, and autonomous regulation---remain theoretically underspecified in the broader literature, with dominant theories offering limited predictive precision about the conditions under which learners sustain or abandon effortful zone transitions \parencite{murayama2026critical}. SCAN addresses this gap
operationally by grounding zone migration in metacognitive control rather than motivational state alone, making the transition criteria explicit and auditable. In human-AI collaboration, the complete absence of metacognitive control introduces a hazardous operational failure mode—documented by \textcite{galindez2026trust}---where users successfully monitor and perceive system anomalies but fail to execute corresponding behavioral changes or strategy shifts, defaulting instead to passive acceptance. This misalignment severely hinders the human operator's capacity to preserve system boundaries. As demonstrated by \textcite{kandul2023human}, when strategic control mechanisms collapse, individuals consistently struggle to map out machine predictability thresholds under real-effort pressures, which directly impairs long-term adaptive task performance.

When executed properly, deliberate \emph{reflection}---as an essential manifestation of retrospective control--activates and restructures prior cognitive frameworks to optimize future decision pathways \parencite{mamede2022reflection}. This reflective mechanism ensures that task distribution is handled as an intentional, self-correcting cognitive policy rather than an automated impulse under structural resource constraints \parencite{corno2001volitional, liu2025metacognitive}. Crucially, this
retrospective process is not purely cognitive: motivational and emotional self-regulation are functionally interwoven with metacognitive control, such that a learner's capacity to revise her zone assignment depends on her ability to regulate both the epistemic judgment and the affective response to perceived failure or over-reliance \parencite{stockinger2026interwoven}.

Taken together, these three components constitute a unified, self-reinforcing regulatory architecture that operationalizes the entire SCAN lifecycle. Metacognitive knowledge frames the initial, anticipatory zone partitioning; monitoring continuously assesses the epistemic validity of human--AI generation in real time; and control retrospectively modifies cognitive policies based on performance outcomes. As control continuously feeds evaluative insights back into the learner's baseline knowledge and monitoring layers (\emph{learning}), the framework evolves from a static taxonomy into a dynamic, closed-loop feedback architecture. Each consecutive interaction progressively refines predictive calibration, minimizes automation vulnerabilities, and supports increasingly proficient, context-dependent human-AI coordination (\cref{fig:illustration}).

\section{Application}

As a human-centric decision-making framework, SCAN can be applied in two primary scenarios. The first one is at the classical supervisor–subordinate relationship \parencite{jablin1979superior}. This aligns closely with ZPD between a learner and a more knowledgeable other. In this context, SCAN operationalizes the decision-making process by helping the supervisor determine which tasks should remain under subordinate's control, which should be supported by, or which can be fully automated by, GenAI. SCAN requires supervisor's metacognitive ability in assessing task complexity, worker's capabilities and limitations, and the suitability of AI assistance. The second one is learner's self-regulated learning perspective \parencite{zimmerman1989social}. Before completing a task, a learner, with her metacognition, evaluates the task at hand, by recognizing if she possesses task specific knowledge, which types of a scaffold is needed, and identifies such task in one of the four sub-zones in SCAN. Once a task is complete, she reflects from the experience, refines her mental models of her and (AI's) capabilities and limitations, and adjusts the scaffolding preferences before feeding back to the next task in the future. In other words, SCAN is the way for "scanning" one's own knowledge space and metacognitively evaluating is GenAI used (Substitute, Complement, Aid) or not (Non-negotiable) with the task at hand. In what follows, we apply SCAN in two common real-life scenarios where GenAI assists as a scaffold: knowledge workers at the workplace \parencite{jarrahi2025generative} and students in education \parencite{wang2024artificial}, which could represent individuals who always have sufficient task-specific knowledge before the rise of GenAI, and those who are currently accumulating task-specific knowledge during the GenAI era.

\subsection{Knowledge workers at the workplace}
Within organizational settings, SCAN functions as an explicit decision protocol that binds task analysis to metacognitive self-monitoring. First, before any delegation, a worker classifies the task with respect to (i) domain-specific competence (``known to learner''), (ii) the anticipated generalizable competence of the model (``known to GenAI''), and (iii) downstream accountability. Using GenAI as scaffold, the \emph{Complement} and \emph{Aid} sub-zones instantiate\emph{collaboration} and \emph{augmentation}: AI contributes breadth (pattern libraries, literature synthesis, exemplar generation), while the human curates goals, constraints, and evaluation criteria, maintaining agency and epistemic accountability \parencite{mollick2024co,wood1976role, gutoreva2026}. In contrast, the \emph{Substitute} sub-zone tempts automation under uncertainty; SCAN therefore requires a pre-commitment to metacognitive checks (calibration prompts, counter-example search, provenance inspection) to mitigate over-offloading and LLM sycophancy \parencite{malmqvist2025sycophancy,ritz2024offloading}. 

\subsection{Students in education setting}
For learners, SCAN operationalizes ZPD by making the \emph{scaffold} explicit and accountable \parencite{Vygotsky1978,wood1976role,yan2024promises}. For a task, instructors require students to (1) divide, and thus classify, each sub-task into S/C/A/N; (2) justify the choice with a brief metacognitive note (why collaboration vs.\ augmentation vs.\ automation?); and (3) provide an audit trail (prompts, drafts, and human revisions) activating student's metacognition. Completing tasks in \emph{Aid} target conceptual bottlenecks where AI hints, exemplars, or Socratic questioning \parencite{favero2024enhancing}, stimulate students' curiosity \parencite{abdelghani2022conversational}, reduce extraneous load while preserving germane processing \parencite{van2024cognitive}. \emph{Complement} tasks cultivate higher-order skills in co-learning \parencite{schoonderwoerd2022design}: when examining GenAI's output after delegation, students compare multiple AI rationales, critique them against rubrics \parencite{suriano2025student}, and integrate disciplinary standards (methods sections, proofs, proper citation) to prevent shallow pattern imitation, while fostering students' autonomy \parencite{zhu2025fostering}. 

The pedagogy emphasizes \emph{metacognitive calibration}, which is what SCAN emphasizes on learning. Students forecast performance before AI use, reflect after, and track shifts across sessions as tasks migrate from S$\rightarrow$A$\rightarrow$C, making progress legible. Where appropriate, lightweight neurophysiological measures (e.g., attention proxies or simple EEG in lab classes) can provide formative biofeedback to time SCAN transitions, but control remains with the learner. Finally, the \emph{Shared Self} lens positions AI not as a ghostwriter but as a transparent cognitive extension under the student’s authorship and responsibility; assessment design (oral defenses, process portfolios, randomized transfer tasks) evaluates \emph{what the student can now do unaided}, not merely the quality of an AI-assisted artifact \parencite{mollick2024co}.

\section{Discussion}
Positing Vygotsky's ZPD and Metacognition, SCAN highlights some of the significant elements in learning with GenAI, and other forms of AI in the future such as Artifical General Intelligence \parencite{goertzel2007artificial}. Under SCAN, a learner can recognize, and thus maintain, an appropriate relationship between herself, more knowledgeable others, and GenAI when completing a task. This is one of the essential foundational skills for achieving long-term mastery as part of lifelong learning in the GenAI era \parencite{gonsalves2024generative}. Secondly, SCAN can serve as \emph{a bridge} between AI literacy and one's metacognitive ability \parencite{bewersdorff2025ai}, especially when a learner monitors and assesses own and GenAI's ability in a task assignment such as problem solving \parencite{flavell1979metacognition, flavell1976metacognitive, cox2005metacognition}. Next, applying SCAN as a human-centric decision making framework means that a learner is always present in the loop (``Human-In-The-Loop"), implying that she can decide whether a task is located in one of the four sub-zones in the framework, as well as agency \parencite{wen2022sense}, control, and associated responsibilities including assessing the assistance from GenAI. This is more prevalent in scenarios where implementing GenAI as agents: user's involvement and GenAI's autonomy fluctuate as the user's role varies over time \parencite{feng2025levels}. Last but not least, SCAN does not eliminate the possibility that there are non-negotiable tasks to humans with other humans, which, in turn, do not have any involvement with GenAI. These tasks are, often, circumstances where AI cannot perform well whereas human excel \parencite{loaiza2024epoch}. 

In what follows, we discuss several relevant aspects of SCAN framework, including cognitive load, cognitive offloading and sycophancy, three decision making modes in human-AI interactions (automation, augmentation and collaboration), future of work such as upskilling and deskilling, and Vygoysky's Internalization. We end this section by discussing limitations of our work, followed by suggesting a few potential paths where future research can extend on SCAN.

\subsection{SCAN's Task Paradigm and Cognitive Load}
The four sub-zones in SCAN's Task Paradigm carry distinct cognitive load profiles on learning and performance. Cognitive Load Theory (CLT) distinguishes three forms of ``load'' \parencite{sweller1998cognitive, paas2003cognitive}: \emph{intrinsic load}, determined by the inherent complexity of the material and the extent to which its elements must be processed simultaneously; \emph{extraneous load}, imposed by how the task is presented or mediated, which interferes with schema acquisition when unnecessary; and \emph{germane load}, the working memory resources actively redirected away from extraneous processing and toward the intrinsic, schema-building work that constitutes genuine learning \parencite{sweller2020cognitive}. Underlying this framework, there is a foundational insight from \parencite{sweller1988cognitive}: a schema is a cognitive structure that allows a learner to recognize a task as belonging to a familiar category and retrieve appropriate responses. It is the primary unit of expertise, and its acquisition competes directly with goal-directed task completion for the same limited processing capacity.

In SCAN, a learner possesses no task-specific schemas at the Substitute sub-zone, implying the intrinsic load of the task already exceeds her current working memory capacity. Delegating to GenAI reduces the felt burden, but at a steep cognitive cost. As she lacks the schemas needed to evaluate GenAI's output critically, extraneous load is paradoxically elevated: she must navigate unfamiliar material without the conceptual anchors that would make that navigation meaningful. The redistributive function of germane load is blocked entirely: with no reduction in extraneous load, no working memory resources can be redirected toward intrinsic schema-building work. The sub-zone therefore carries the highest cognitive risk: it feels efficient while quietly arresting development \parencite{stadler2024cognitive}.

A learner's load profile inverts relative to Substitute at the Complement sub-zone. Intrinsic load is manageable as the learner now possesses task-specific schemas; extraneous load is low because she can evaluate GenAI's output efficiently using those schemas. As both loads contained well within limits of working memory, the germane load operates at its fullest: resources no longer needed for navigating the task itself are directed toward higher-order schema extension such as quality assurance, strategic refinement, and transfer to adjacent domains \parencite{sweller2019cognitive}. Complement is, as SCAN suggests, where human-AI collaboration delivers its clearest productivity dividend without sacrificing epistemic ownership.

A learner's load profile becomes generative at the Aid sub-zone. Intrinsic load remains non-trivial as the task sits at the learner's ZPD, but but GenAI's scaffolding reduces extraneous load by providing structure, exemplars, or targeted hints without resolving the task entirely. This reduction in extraneous load activates germane processing: working memory resources are now freed to be redirected toward the intrinsic schema-building work the learner must still perform \parencite{sweller2019cognitive}. Aid is SCAN's pedagogically optimal sub-zone: the configuration that most directly instantiates CLT's prescriptive logic of reducing extraneous load in order to maximize resources available for schema construction \parencite{sweller1998cognitive, paas2003cognitive}.

Non-negotiable occupies a categorically different position in this load landscape. Intrinsic load is high, as tasks in this sub-zone involve tacit knowledge, value judgments, identity-relevant deliberation, or normatively complex material whose elements must be processed simultaneously without the simplification that schemas typically provide. Extraneous load is also present, but its source is fundamentally different from Substitute: : rather than arising from navigating GenAI output without anchoring schemas, it arises from the relational and interpretive complexity of human interaction itself--the ambiguity of mentorship, the contextual judgment required in supervision, the negotiation of contested evaluative criteria \parencite{morewedge2022preference}. Germane load is activated, as human experts can progressively reduce that interpretive uncertainty, freeing working memory resources for the value-laden schema formation that is Non-negotiable's developmental purpose. 

These load profiles are, under SCAN, \emph{dynamic} over time (\cref{tab:cognitive_load_profiles}). As metacognition drives the sub-zones migration described in the subsequent sections, the same task type travels through qualitatively different load landscapes over time. A task that initially saturates working memory at Substitute gradually becomes Aid as schemas consolidate, and eventually Complement as task-specific knowledge becomes fluent. At each transition, the relationship between intrinsic and extraneous load shifts: intrinsic load decreases as schemas reduce the number of elements requiring simultaneous processing, which, in turn, reduces total load and further amplifies the resources available for germane redistribution \parencite{sweller2019cognitive}. In short, SCAN provides a principled account of how effective scaffolding should evolve: shifting the locus of cognitive work from GenAI to the learner, as schemas form and working memory capacity is progressively freed for the intrinsic processing through which expertise is built over time.

\begin{table}[htbp]
    \centering
    \caption{Cognitive Load Profiles Across SCAN's Four Sub-zones}
    \label{tab:cognitive_load_profiles}
    \renewcommand{\arraystretch}{1.3} 
    \begin{tabularx}{\textwidth}{c C C C} 
        \toprule
        \textbf{Sub-zone} & \textbf{Intrinsic Load} & \textbf{Extraneous Load} & \textbf{Germane Load}\\
        \midrule
        \textbf{S} & High & High & Blocked\\
        \textbf{C} & Low & Low & Maximized at higher order\\
        \textbf{A} & Moderate & Reduced & Activated\\
        \textbf{N} & High & Moderate & Activated through dialogue\\
        \bottomrule
    \end{tabularx}
\end{table}

\subsection{Cognitive Offloading and Sycophancy}
A learner is prone to cognitive offloading \parencite{shukla2025skilling} and sycophancy \parencite{sharma2023sycophancy} when interacting with GenAI. \emph{Cognitive Offloading} denotes the reduction of a learner's mental effort a particular task by using external tools like GenAI \parencite{ritz2024offloading}. \emph{Sycophancy} refers to GenAI's explicit and implicit agreement on a learner's prompt over evidence-based results, and thus reinforcing her preferences and beliefs, which are incorrect \parencite{malmqvist2025sycophancy, cheng2025elephantmeasuringunderstandingsocial}. SCAN suggests that their degrees on a learner vary across three sub-zones due to different levels of task-specific knowledge one possesses about a task (\cref{tab:sub-zones}). For instance, if a learner's task is at the Substitute rather than the Aid, she is more prone to both cognitive offloading (as she lacks task-specific knowledge for task completion) and sycophancy (she could not recognize, and thus verify, whether GenAI provides an evidence-based response). If she is working on a task at the Complement, her task-specific knowledge, with her metacognition, mitigates cognitive offloading and sycophancy to a large extent, by monitoring, and challenging GenAI's output via counter-arguments. 

\begin{table}[htbp]
    \centering
    \caption{Analysis of SCAN sub-zones: cognitive offloading and sycophancy}
    \label{tab:sub-zones}
    \renewcommand{\arraystretch}{1.3} 
    \begin{tabularx}{\textwidth}{c C C C}
        \toprule
        \textbf{Sub-zone} & \textbf{Level of task-specific knowledge} & \textbf{Level of Cognitive Offloading} & \textbf{Proneness to Sycophancy} \\
        \midrule
        C & Sufficient & Low & Low \\
        A & Some but insufficient & Medium & Medium \\
        S & None & High & High \\
        \bottomrule
    \end{tabularx}
\end{table}

As SCAN embeds the self-sustaining cycle of a learner's metacognition (real-time evaluation, reflection, and learning) when completing a task with GenAI (S, A, C sub-zones), an interesting question is how her cognitive cost would be influenced due to cognitive offloading in the long term \parencite{inie2025cognitive}, which current empirical studies could capture such impact in the short term. We think that SCAN could offer a possible direction to mitigate the long-term cognitive offloading from a learner's perspective, by (1) identifying which sub-zone a task belongs to with GenAI as a scaffold, as well as (2) iteratively reflecting from the completed task, refining her mental models of her and (AI's) capabilities and limitations, and adjusting the scaffolding preferences before feeding back to the next task. Apart from that, as task-specific knowledge can be accumulated via learner's self-sustaining cycle of metacognition, GenAI's sycophantic nature would, perhaps, have a less impact on learner herself over time.

\subsection{Three Human-AI Decision Making Modes}

There are three main decision-making modes when human interact with GenAI: automation, augmentation, and collaboration \parencite{jiang2024human}. We distinguish \emph{collaboration} and \emph{augmentation} from \emph{automation} by locus of control and epistemic responsibility rather than by tool sophistication. First, \textbf{Automation} denotes the transfer of initiative and intermediate decision rights to an agentic system for well-specified subroutines whose failure modes are bounded and recoverable, and thus where human “trust” is built towards AI \parencite{glikson2020human, schaefer2016meta}. Next, \textbf{Augmentation} denotes the enhancement of human capabilities by leveraging GenAI's capabilities, which performs better than human working alone in a task \parencite{vaccaro2024combinations}. Finally, \textbf{Collaboration} denotes deliberately maintained human agency over goal specification, constraint setting, evaluation criteria, and final endorsement, with GenAI providing breadth (search, recall, exemplar generation), structure (drafting, decomposition), or challenge (counter-arguments, red-teaming) \parencite{berretta2023defining}. 

 In the SCAN framing where a human interacts with GenAI, \emph{Substitute} is the sub-zone that tempts automation; \emph{Aid} instantiates augmentation; and \emph{Complement} represents collaboration. The design imperative is to ensure that movement into S does not silently degrade metacognition, calibration, or authorship. We, therefore, advocate a \emph{reversible delegation} principle: any transition from collaboration or augmentation to automation requires (i) a capability threshold test (held-out tasks, shifting distributions), (ii) provenance guarantees (traceable inputs/edits), and (iii) a human-in-the-loop veto with cost-aware rollback. This aligns with ZPD’s scaffolding logic \parencite{Vygotsky1978,wood1976role} and contemporary guidance on responsible task separation for knowledge work \parencite{mollick2024co}.

Operationally, SCAN supplies guardrails that keep automation honest. First, \emph{risk-weighted routing}: tasks with high normative content, tacit knowledge, contested objectives, or complex externalities remain in the \emph{Non-negotiable}—human-led regardless of model capability \parencite{feng2025levels}. Second, \emph{sycophancy hardening}: before allowing S-level delegation, systems must pass contradiction probes and stance-invariance tests; when models over-agree with user priors, delegation is rejected or downgraded to augmentation \parencite{malmqvist2025sycophancy}. Third, \emph{offloading budgets}: organizations set quantitative ceilings on S-time per project and require post-hoc variance analysis on human edits, curbing the well-documented drift toward excessive cognitive offloading \parencite{ritz2024offloading}. Next, \emph{learning-preserving pipelines}: even when automation is justified (stable, repetitive, safety-bounded subroutines), the pipeline periodically re-inserts \emph{augmentation or collaboration checkpoints}—short, reflective compare-and-contrast episodes that protect skill retention and situational awareness, provided that human leads, and is, in the loop \parencite{van2024cognitive}. Finally, \emph{Automation-Augmentation Paradox}: SCAN sheds lights on understanding the interdependent nature between automation and augmentation \parencite{agrawal2023we}, as well as collaboration, which can be seen as a continuum one's metacognitive ability over time \textcite{raisch2021artificial}. 

\subsection{Upskilling and Deskilling}
SCAN sheds some lights on the future of work \parencite{frey2017future, khaokaew2022imagining}, by offering explanations on when, and how, upskilling and deskilling occur when a learner works with GenAI. After completing a task, a learner, with her metacognition, feeds back what she learned from a current task into real-time evaluation of a new task. Such self-sustaining cycle of real-time evaluation, reflection, and learning in her metacognition leads to an accumulation of learner's task specific knowledge when the same type of task is identified differently over time. In what follows, we provide a basic mathematical illustration of how such changes across three sub-zones in SCAN lead to upskilling.

We denote that a learner's task specific knowledge can be divided into three ``levels'' in SCAN: \emph{beginner}, \emph{intermediate}, and \emph{expert}, corresponding to \emph{none}, \emph{some but insufficient}, and \emph{sufficient} task specific knowledge as we described when introducing each sub-zone in SCAN. Next, we formulate SCAN by adding the time dimension $t$, illustrating how learning with GenAI occurs via tasks completion over time.  

\begin{equation}
    KL_{Beginner}=\frac{S}{S+C+A+N}
\end{equation}

\begin{equation}
    KL_{Intermediate}=\frac{A}{S+C+A+N}
\end{equation}

\begin{equation}
    KL_{Expert}=\frac{C}{S+C+A+N}
\end{equation}

As for the upskilling, a \emph{beginner} learner becomes an \emph{intermediate} learner when she has some but insufficient task specific knowledge at $t+1$, provided that she has none task specific knowledge at $t$, implying

\begin{equation}
    A_{t+1} - S_{t+1} > 0
\end{equation}

Similarly, an \emph{intermediate} learner becomes an \emph{expert} learner when when she has sufficient task specific knowledge at $t+1$, provided that she has some task specific knowledge at $t$,

\begin{equation}
    C_{t+1} - A_{t+1} > 0
\end{equation}

In other words, SCAN treats learning with GenAI as systematic migration of tasks across sub-zones over time: S$\rightarrow$A$\rightarrow$C \footnote{We note that while we emphasize on learning with GenAI in this section, N provides a normative boundary that resists erosion by capability creep, which we discuss briefly in the following section.}. Pedagogically and organizationally, the core is \emph{metacognitive instrumentation}. Each task instance is logged with (a) the learner’s pre-use forecast, (b) the chosen SCAN sub-zone and justification, (c) AI interaction traces, (d) post-task calibration (confidence vs.\ accuracy), and (e) transfer probes to adjacent tasks. These traces surface which knowledge components are genuinely internalized versus merely performed by the model \parencite{yan2024promises,van2024cognitive}. Over time, learners can aim to reduce reliance on S by converting opaque routines into explainable sub-skills inside A, then into independently executable patterns in C. The quantitative indices already defined in this paper (e.g., $KL_{\text{Beginner}}{=}\frac{S}{S{+}C{+}A{+}N}$, $KL_{\text{Intermediate}}{=}\frac{A}{S{+}C{+}A{+}N}$, $KL_{\text{Expert}}{=}\frac{C}{S{+}C{+}A{+}N}$) provide a compact summary of trajectory; improvement targets are formalized as $(A_{t+1}-S_{t+1})>0$ and $(C_{t+1}-A_{t+1})>0$ transitions, coupled with gains in calibration and far transfer.

In SCAN, upskilling proceeds by deliberately shifting recurrent tasks from \emph{Substitute} to \emph{Aid} to \emph{Complement} via spaced practice and reflective comparison of human vs.\ AI rationales (error tagging, ``explain-back'' prompts). In high-load or safety-critical contexts, optional BCI augmentation can act as a closed-loop \emph{metacognitive sentinel}: when neural or physiological markers indicate overload or shallow processing, the system triggers SCAN nudges (e.g., from S to A/C) without seizing control. This preserves autonomy and aligns with the \emph{Shared Self} perspective, wherein AI systems are incorporated as transparent extensions of the agent’s problem-solving apparatus while responsibility and identity remain stably anchored in the human \parencite{gutoreva2024sharing}. Anticipating stronger model capabilities (AGI-adjacent scenarios), SCAN’s \emph{Non-negotiable} sub-zone codifies domains that remain human-led, with more knowledgeable others, due to normative, legal, or tacit-knowledge constraints (e.g., value judgments, personnel decisions, consent and accountability), ensuring that increased capability does not silently erode agency.

Two mechanisms accelerate upskilling without sacrificing learner's agency. First, \emph{explain-back practice}: whenever AI is used in A/C, a learner must reconstruct the reasoning, label assumptions, and generate a minimal, human-authored reference solution; evaluation privileges the explain-back, not the AI artifact. Second, \emph{difficulty titration}: instruction dynamically adjusts task complexity to sit inside A (productive struggle), not S (overload) nor trivial C (underload), closely mirroring ZPD’s sweet spot for growth \parencite{Vygotsky1978,wood1976role}. In high-load environments, optional BCI adds a closed-loop \emph{metacognitive sentinel}: physiological markers of overload trigger nudges to switch from S to A (ask for hints, not answers) or to pause for reflection, while markers of mindless fluency in C trigger prompts to escalate difficulty. Throughout, the \emph{Shared Self} lens frames AI as a transparent extension of the learner’s cognitive apparatus rather than a surrogate author, maintaining stable responsibility and identity as the competence grows \parencite{gutoreva2024sharing}. The result is not merely faster task completion, but measurable, durable capability formation: fewer S-classifications, denser C-competence, higher calibration, and demonstrable transfer to novel tasks—even when the AI is switched off.

On the other hand, in SCAN, deskilling \parencite{ferdman2025ai, shukla2025skilling} occurs when tasks are identified from Complement to Aid. In what follows, we show how both upskilling and deskilling occur in the same (type of) task at three different points of time ($t$, $t+1$, $t+2$) for a simple illustration. Upskilling occurs in a task when a learner has some task specific knowledge, and can complete with GenAI assistance (e.g., $A_{t} - S_{t} > 0$). Over time, a learner would, with her metacongitive ability, equip more relevant task-specific knowledge from her previous task completion experience. Thus, when she encounters the same or a similar task, she might identify it as a task at the Complement, and thus prefer to delegate GenAI for task completion (e.g., $C_{t+1} - A_{t+1} > 0$). If her preferences of delegating similar tasks to GenAI were consistent over time, skill erosion could occur as she lacks practice while over-relying on GenAI outputs, leading to knowledge detention and deskilling \parencite{barcaui2025chatgpt}. Thus, this (type of) task could likely fall back in the Aid, as she still has some task-specific knowledge (e.g., $C_{t+2} - A_{t+2} < 0$). As a result, one might expect that identification of such (type of) tasks might oscillate between the Aid and Complement zones in the long term, aligning with growing research investigating a potential \emph{upskilling-deskilling paradox} in human-AI interactions \parencite{shukla2025skilling, krook2025autonomy}. 

\subsection{How SCAN Accounts for Both Human-Human and Human-AI Learning}

As an extension of ZPD in the GenAI era, SCAN is designed to account for two distinct and complementary learning trajectories simultaneously. The first is \emph{human-human learning}, in which a learner completes tasks within Non-negotiable---the sub-zone that preserves Vygotsky's ZPD in its original form, requiring the guidance of a human MKO rather than GenAI. The second is \emph{human-AI learning}, in which a learner navigates from Substitute through Aid to Complement, progressively developing task-specific knowledge and agency in collaboration with GenAI. Understanding how each trajectory operates requires examining how the internalization arc, which is the mechanism Vygotsky identified as central to human development, functions differently across them.

The internalization arc describes a directional developmental process in Vygotsky's original formulation. Transformation from social to individual, from other-regulation to shared regulation to self-regulation leads to developmental, cognitive growth \parencite{wertsch1988vygotsky}. In Non-negotiable, the internalization arc operates as Vygotsky originally described. The human MKO scaffolds, models professional judgment, and gradually withdraws support as the learner internalizes the relevant capacities such as cognitive \parencite{collins1989cognitive}, behavioral \parencite{bandura1977social}, and social \parencite{lave1991situated} until independent performance is achieved. What the learner acquires are both task-specific knowledge, and more importantly, tacit, situated expertise that cannot be codified or transmitted through general knowledge \parencite{polanyi1966tacit}. The latter is the kind of knowing-how that \textcite{dreyfus1986mind} argue is structurally beyond what rule-based computational systems can replicate. External manifestations of the mediatory link progressively diminish, and the MKO's guidance is incorporated into the learner's internal constitution as professional judgment and identity \parencite{damianova2011rereading}. The endpoint is \emph{complete internalization}: the scaffold disappears, and the learner becomes capable of serving as an MKO for others, completing the social learning cycle \parencite{lave1991situated}. 

In Substitute, Aid, and Complement, the internalization arc operates differently in nature. As a learner moves from S to A to C, task-specific knowledge is accumulated, and agency is reclaimed progressively. However, neither marker of Vygotsky's Internalization is fully met: GenAI never disappears from the external plane, and its operations are not incorporated into the learner's internal constitution. It is a different developmental process that which Vygotsky's framework offers no account as GenAI did not exist. We propose that this process is better understood as the progressive integration of AI as part of the learner's extended self: GenAI increasingly participates in attention allocation, reasoning, synthesis, and decision-making, shaping the very cognitive processes through which the learner forms beliefs, makes decisions, and constitutes their sense of self \parencite{gutoreva2026}. In this view, the learner and GenAI form a \emph{symbiotic cognitive unit}: cognition is realized dynamically through interactions with AI systems. We term this developmental process the \emph{Symbiosis Trajectory}, in contrast to Vygotsky's classical \emph{Internalization Trajectory} in which the scaffold fades and the learner arrives at autonomous self-regulation. This integration orients the human-GenAI unit toward \emph{hybrid intelligence}: the ability to achieve complex goals by combining human and artificial intelligence in ways that neither could accomplish separately \parencite{hamilton2019human, steyvers2022bayesian, gonzalez2025cognitive}. In short, the Symbiosis Trajectory is how learners develop the collaborative competence to work with GenAI throughout a lifetime.

Despite their different logics, both trajectories share two common normative dimensions that increase across both arcs, as the learner moves toward self-regulation (Table~\ref{tab:agency_levels}). The first dimension is \emph{epistemic agency}, which is the learner's capacity to act as a knowing subject, evaluating and questioning knowledge actively \parencite{nieminen2024epistemic, nieminen2025active}. It is followed by \emph{epistemic responsibility}, which is the obligations that accompany that capacity in high-stakes domains \parencite{code1987epistemic}. In N, this culminates in the complete independence and accountability of the expert practitioner. In S to A to C, it culminates in the metacognitive sophistication of the proficient human-AI collaborator---one who directs, evaluates, and takes responsibility for GenAI's outputs rather than deferring to them. The goal of S to A to C is to integrate GenAI into one's cognitive practice as a genuine extension of one's knowing self \parencite{gutoreva2026}.

\begin{table}[ht]
    \centering
    \caption{Levels of Agency and Responsibility in Decision-Making Modes}
    \label{tab:agency_levels}
    \begin{tabularx}{\textwidth}{@{} CCCCC @{}}
        \toprule
        \textbf{Zone} & 
        \makecell{\textbf{Decision-making} \\ \textbf{mode}} & 
        \makecell{\textbf{Internalization} \\ \textbf{stage}} & 
        \makecell{\textbf{Epistemic} \\ \textbf{agency}} & 
        \makecell{\textbf{Epistemic} \\ \textbf{responsibility}} \\ 
        \midrule
        Substitute           & Automation   & Other-regulation  & Low    & Low    \\
        Aid / Non-Negotiable & Augmentation & Shared regulation & Medium & Medium \\
        Complement           & Collaboration & Self-regulation   & High   & High   \\ 
        \bottomrule
    \end{tabularx}
\end{table}

This architecture responds directly to concerns that GenAI may gradually erode distinctly human developmental processes \parencite{fernandes2026ai,inie2025cognitive, salatino2025influence}. By preserving Non-negotiable as a theoretically grounded zone in which complete internalization remains the endpoint, SCAN ensures that the human-human developmental arc Vygotsky described is actively protected within it. Emerging evidence further suggests that well-designed human-AI interaction can itself scaffold metacognitive development---prompting learners to surface implicit assumptions, and engage in deliberate self-monitoring that unassisted study rarely elicits \parencite{jarvela2023human}. SCAN's contribution is to specify the conditions under which each trajectory is appropriate, ensuring that both are available to the learner who needs them.

\subsection{Limitations}
We acknowledge several limitations when applying SCAN Framework. First, the SCAN framework requires metacognitive capacity that novice learners may not yet possess, which is crucial for classifying tasks into SCAN four sub-zones. This could be mitigated when more knowledgeable others such as educators introduce the framework, and thus supervise task classification activities with them. Secondly, task composition in practice is complex, and thus difficult. This could be resolved by linking the task objectives and outcomes. Third, SCAN framework is not applicable when external constraints such as time pressure and decision contexts are constantly changing (e.g., dynamic decision making). However, SCAN could be used as a reflective tool (part of metacognition), which in turn, adapting SCAN framework as a habit in upcoming AI-assisted decision making scenario. Finally, assessing SCAN's effectiveness is complicated, especially in education setting where standard rubics and assessment procedures apply to all students. As SCAN is an extension of Vygotsky's ZPD, assessment tools for ZPD also applies to assessing SCAN's effectiveness, e.g.,``dynamic assessmen'' that empirically measures the gap between what a learner can do independently and assisted performance \parencite{allal2000assessment, baek2003effect}, and quasi-experimental pre/post designs that comparing ZPD-assisted conditions against independent practice to quantify learning gains \parencite{zbainos2019investigating, nassaji2000vygotskian}.

\subsection{Future Research}
With the foundation SCAN builds, we recognize several paths future research can investigate, and thus deepen our understanding on building, and sustaining an effective relationship between human and GenAI. First, one can formulate the SCAN mathematically, and thus strengthen its testability. Some research investigated ZPD mathematically, known as the ``gray area'' \parencite{chounta2017grey}, which we believe can be a great start. Also, this allows us to explore how does Non-negotiable plays a role in knowledge transfer and accumulation for human-human learning in the GenAI era. Secondly, we see the potential of incorporating pre-existing works of modeling knowledge and research of human cognition, such as \textcite{corbett1994knowledge, anderson1999practice} into SCAN, which we presented a basic illustration of how task-specific knowledge is formed, stored, and accumulated as a learner completes more tasks over time. This incorporation facilitates our understanding on ongoing issues related to the future of work, such as upskilling and deskilling. We view this offers a practical pathway of enhancing the former and mitigating the latter. Apart from expanding the SCAN framework, we suggest an empirical investigation that compares both human-centric and AI-centric approach, like \textcite{fugener2025roles}, in terms of task assignments. As the decision-making modes in human-AI interactions are task-dependent, experiments foster our understanding of whether a human-centric or AI-centric approach is preferred for better performance in a particular task type, both in the short and long term. Last but not least, SCAN has shed some lights on three ongoing discussions regarding human-AI interactions, where future research can apply SCAN for further investigations. First, the Aid and Non-negotiable mirror \emph{Algorithmic Appreciation} \parencite{logg2019algorithm} and \emph{Algorithmic Aversion} \parencite{dietvorst2015algorithm} respectively. SCAN suggests task-dependence as a possible explanation for such ongoing discussion, which future research can investigate further, and deepen our understanding on where the boundary lies. Apart from that, SCAN offers a plausible explanation for the Augmentation-Automation paradox \parencite{raisch2021artificial}: a learner's accumulation of task-specific knowledge with her metacognition ability when interacting with GenAI over time. Next, we discuss a potential paradox that lies between upskilling and deskilling, which could be due to a shift of user's role from production to evaluation. SCAN suggests such shift is due to the different levels of task-specific knowledge a learner possesses between Aid and Complement. Future research can empirically investigate these two paradoxes in various types of task, such as creative and decision tasks \parencite{vaccaro2024combinations}.

\section{Conclusion}
We introduce SCAN as a human-centric, decision-making framework that facilitates effective task allocation with GenAI, which can, in turn, increase an individual's GenAI literacy by ``scanning'' the AI-human interaction in task completions. We propose that SCAN offers a clear direction for learners, regardless of possessing sufficient task-specific knowledge or not, complete a task with GenAI assistance effectively in the long term. Built upon Vygotsky's ZPD and Metacognition, we see SCAN as a plausible decision-making framework that applies to other advanced forms of AI in the future, such as AGI, where human (lifelong) learning is, unsurprisingly and undoubtedly, necessary.

\printbibliography  

\end{document}